\documentclass[12pt]{article}

\def\comp{{\rm C}\llap{\vrule height7.1pt width1pt depth-.4pt\phantom t}}

\def\square{\kern1pt\vbox{\hrule height 1.2pt\hbox{\vrule width 1.2pt\hskip 3pt
   \vbox{\vskip 6pt}\hskip 3pt\vrule width 0.6pt}\hrule height 0.6pt}\kern1pt}

\begin{document}
\begin{titlepage}
\begin{flushright}
UFIFT-QG-06-04 \\ gr-qc/0608049
\end{flushright}
\vspace{1cm}
\begin{center}
\textbf{One Loop Corrected Mode Functions for SQED during Inflation}
\end{center}
\vskip 1cm
\begin{center}
E. O. Kahya$^{\dagger}$ and R. P. Woodard$^{\ddagger}$
\end{center}
\begin{center}
\textit{Department of Physics \\ University of Florida \\
Gainesville, FL 32611 USA}
\end{center}
\vskip 1cm
\begin{center}
ABSTRACT
\end{center}
We solve the one loop effective scalar field equations for spatial plane
waves in massless, minimally coupled scalar quantum electrodynamics 
on a locally de Sitter background. The computation is done in two different
gauges: a non-de Sitter invariant analogue of Feynman gauge, and in the
de Sitter invariant, Lorentz gauge. In each case our result is that the
finite part of the conformal counterterm can be chosen so that the mode
functions experience no significant one loop corrections at late times.
This is in perfect agreement with a recent, all orders stochastic prediction.

\begin{flushleft}
\end{flushleft}
\vspace{.4cm}
\begin{flushleft}
$^{\dagger}$ e-mail: emre@phys.ufl.edu \\
$^{\ddagger}$ e-mail: woodard@phys.ufl.edu
\end{flushleft}
\end{titlepage}

\section{Introduction}

Gravitons and massless, minimally coupled (MMC) scalars are unique in 
possessing zero mass without classical conformal invariance. This allows 
them to mediate vastly enhanced quantum effects during inflation \cite{RPW1}. 
Many examples have been studied over the history of inflation, starting 
with the first work on scalar \cite{MC} and tensor \cite{AAS1} density 
perturbations. Those were tree order analyses. More recently, there
have been analyses of loop corrections to density perturbations 
\cite{SW1,BVS1,BVS2,MS} and to similar fixed-momentum correlators \cite{SW2}.

A reasonable paradigm for much of inflation is de Sitter background and a 
wide variety of enhanced quantum loop effects have been studied on this
background in many different theories. In pure quantum gravity the one loop 
self-energy \cite{TW1} and the two loop expectation value of the metric 
\cite{TW0,TW2} have been calculated. For a MMC scalar with a quartic self 
interaction the expectation value of the stress tensor \cite{OW1,OW2} and 
the self-mass-squared \cite{BOW} have both been computed at two loop order. 
In scalar quantum electrodynamics (SQED) the vacuum polarization 
\cite{PTW1,PTW2} and the scalar self-mass-squared \cite{KW} were evaluated
at one loop order, and the expectation value of two scalar bilinears have 
been obtained at two loop order \cite{PTsW1}. In Yukawa theory the one loop 
fermion self-energy \cite{PW1,GP} and the scalar self-mass-squared \cite{DW} 
have both been evaluated at one loop order. A leading logarithm resummation 
has also been obtained for Yukawa, and checked against an explicit two loop 
computation \cite{MW1}. And the one loop fermion self-energy has been 
calculated in Dirac + Einstein \cite{MW2}.

Of course density perturbations are directly observable, as are the 
expectation values of operators such as the stress tensor. However, an
additional step is necessary in order to infer physics from a one particle 
irreducible (1PI) function such as the vacuum polarization $[\mbox{}^{\mu}
\Pi^{\nu}](x;x')$, the fermion self-energy $[\mbox{}_i\Sigma_j](x;x')$, or
the scalar self-mass-squared $M^2(x;x')$. These 1PI functions correct the 
classical linearized field equations as follows,
\begin{eqnarray}
\partial_{\nu} \Bigl(\sqrt{-g} g^{\nu\rho} g^{\mu\sigma} \mathcal{F}_{\rho
\sigma}(x)\Bigr) + \int d^4x' \Bigl[\mbox{}^{\mu} \Pi^{\nu}\Bigr](x;x')
\mathcal{A}_{\nu}(x') = 0 \; , \\
\sqrt{-g} e^{\mu}_{~a} \gamma^a_{ij} \Bigl(\partial_{\mu} + \frac{i}2
A_{\mu bc} J^{bc}\Bigr)_{jk} \Psi_k(x) - \int d^4x' \Bigl[\mbox{}_i \Sigma_j
\Bigr](x;x') \Psi_j(x') = 0 \; , \\
\partial_{\mu} \Bigl(\sqrt{-g} g^{\mu\nu} \partial_{\nu} \Phi(x) \Bigr)
- \int d^4x' M^2(x;x') \Phi(x') = 0 \; .
\end{eqnarray}
One can infer physics from the quantum-corrected mode functions.

Doing this for photons in SQED \cite{TW2}, and for fermions in Yukawa 
\cite{TW1}, leads to the surprising result that the inflationary production 
of scalars endows these particles with mass. For fermions in Dirac + Einstein 
one finds that the field strength grows by an amount which eventually
becomes nonperturbatively large \cite{MW3}. On the other hand, the result
for the scalars of Yukawa theory is that the finite part of the conformal 
counterterm can be chosen so that there are no significant late time 
corrections at one loop order \cite{DW}. The purpose of this paper is to 
demonstrate that the same result applies for the scalar mode functions of
SQED.

We begin in section 2 by defining the effective mode equation and the sense
in which we solve it. We also explain the relation between the
$\comp$-number solutions we shall find and the quantum operators of SQED. Of
course the self-mass-squared of SQED is gauge dependent, and has been computed
in two different gauges. Section 3 solves for the mode functions in a non-de
Sitter invariant version of Feynman gauge \cite{RPW2,KW}; section 4 does the 
same for the de Sitter invariant Lorentz gauge \cite{AJ,TW3}. In both cases 
our result is that the finite part of the conformal counterterm can be chosen
so that there are no significant one loop corrections at late times. Section
5 summarizes our work and discusses its implications.

\section{The Effective Mode Equation}

It turns out that the same operator formalism gives rise to many different
effective field equations in quantum field theory. The purpose of this section 
is to specify both the particular ones we are solving and the sense in which 
we shall solve them. We begin by recalling the operator formalism of SQED 
\cite{KW,PTsW1}. We then parameterize general quantum corrected mode functions
and contrast the parameter selections appropriate to flat space scattering 
problems with the choices appropriate to cosmology. A brief review is given
of the Schwinger-Keldysh formalism whose linearized effective field equations
give the quantum corrected mode functions for cosmology. We close with a
discussion of the limitations on our knowledge and what they imply about the
sense in which we should solve the effective mode equation.

\subsection{Relation to Fundamental Operators}

We work on the nondynamical metric background of de Sitter in conformal 
coordinates with spacelike signature,
\begin{equation}
ds^2 \equiv g_{\mu\nu} dx^{\mu} dx^{\nu} = a^2(\eta) \Bigl(-d\eta^2 + d\vec{x}
\!\cdot\! d\vec{x}\Bigr) \qquad {\rm where} \qquad a(\eta) = -\frac1{H \eta} 
\; . \label{gmn}
\end{equation}
In other words, $g_{\mu\nu} = a^2 \eta_{\mu\nu}$, where $\eta_{\mu\nu}$ is
the Minkowski metric. When expressed in terms of renormalized fields and
couplings, the Lagrangian of SQED takes the form,
\begin{eqnarray}
\lefteqn{\mathcal{L} = \frac14 F_{\rho\sigma} F_{\mu\nu} g^{\rho\mu} 
g^{\sigma\nu} \sqrt{-g} - \Bigl(\partial_{\mu} \!-\! i e A_{\mu}\Bigr) 
\varphi^* \Bigl(\partial_{\nu} \!+\! i e A_{\nu}\Bigr) \varphi g^{\mu\nu}
\sqrt{-g} } \nonumber \\
& & \hspace{1cm} -\delta Z_2 \Bigl(\partial_{\mu} \!-\! i e A_{\mu}\Bigr) 
\varphi^* \Bigl(\partial_{\nu} \!+\! i e A_{\nu}\Bigr) \varphi g^{\mu\nu} 
\sqrt{-g} - \delta \xi \varphi^* \varphi R \sqrt{-g} \nonumber \\
& & \hspace{4cm} -\frac14 \delta Z_3 F_{\rho\sigma} F_{\mu\nu} g^{\rho\mu} 
g^{\sigma\nu} \sqrt{-g} - \frac14 \delta \lambda (\varphi^* \varphi)^2 
\sqrt{-g} \; . \qquad
\end{eqnarray}
We have chosen to study the exactly massless version of the theory in
which the renormalized values of the conformal and quartic couplings 
vanish. Hence the parameters have the following expansions in terms of 
the charge $e$,
\begin{eqnarray}
\delta \xi = e^2 \delta \xi_2 + e^4 \delta \xi_4 + \ldots \qquad & , & \qquad
\delta \lambda = e^4 \delta \lambda_4 + \ldots \; , \\
\delta Z_2 = e^2 \delta Z_{2,2} + e^4 \delta Z_{2,4} + \ldots \qquad & , &
\qquad \delta Z_3 = e^2 \delta Z_{3,2} + e^4 \delta Z_{3,4} \; .
\end{eqnarray}

The corrected mode functions for which we will solve are related to the
fundamental operators through a number of choices which require extensive
explanations. To fix notation for these explanations we begin by stating
the relations,
\begin{eqnarray}
\Phi(x;\vec{k}) & = & \Bigl\langle \Psi_f \Bigl\vert \Bigl[\varphi(x),\alpha^{
\dagger}(\vec{k})\Bigr] \Bigr\vert \Psi_i\Bigr\rangle \; , \label{calphi} \\
\mathcal{A}_{\mu}(x;\vec{k},\lambda) & = & \Bigl\langle \Psi_f \Bigl\vert 
\Bigl[A_{\mu}(x),\gamma^{\dagger}(\vec{k},\lambda)\Bigr] \Bigr\vert 
\Psi_i\Bigr\rangle \; . \label{calA} \qquad
\end{eqnarray}
The quantities to be explained are the states $\vert \Psi_i\rangle$ and
$\vert \Psi_f\rangle$, and the free creation operators $\alpha^{\dagger}(
\vec{k})$ and $\gamma^{\dagger}(\vec{k},\lambda)$. 

Flat space scattering problems correspond to taking $\vert \Psi_i\rangle$ to be
the state whose wave functional is free vacuum at asymptotically early times.
One also chooses $\vert \Psi_f\rangle$ to be the state whose wave functional 
is free vacuum at asymptotically late times. Although these choices have
great physical interest for flat space scattering problems, they have little
relevance for cosmology. In cosmology the universe often begins at a finite
time, and it evolves to some unknown state in the asymptotic future. Persisting
with the choices of flat space scattering theory would result in acausal
effective field equations which are dominated by the imperative of forcing
the universe to approach free vacuum. The fact that we are looking at 
matrix elements, rather than expectation values, would also have the curious
consequence of giving complex results for the matrix elements of Hermitian
operators.

In cosmology it is better to imagine releasing the universe from a prepared
state at some finite time, and then watching it evolve as it will. This
corresponds to the choice $\vert \Psi_f \rangle = \vert \Psi_i \rangle$. We
additionally assume that both are free vacuum at $\eta \!=\! \eta_i$.

So much for the states in relations (\ref{calphi}-\ref{calA}). To understand
the free creation and annihilation operators one integrates the invariant
field equations of SQED,
\begin{eqnarray}
D_{\mu} \Bigl(\sqrt{-g} g^{\mu\nu} D_{\nu} \varphi\Bigr) - \frac{\delta \xi
\sqrt{-g} R \varphi}{1 \!+\! \delta Z_2} - \frac{\delta \lambda \sqrt{-g}
\varphi^* \varphi^2}{2 (1 \!+\! \delta Z_2)} = 0 \; , \label{phieqn} \\
\partial_{\nu} \Bigl(\sqrt{-g} g^{\nu\rho} g^{\mu\sigma} F_{\rho \sigma}\Bigr)
+ \frac{i e \sqrt{-g} g^{\mu\nu}}{1 \!+\! \delta Z_3} \Bigl(\varphi^* D_{\mu}
\varphi \!-\! (D_{\nu} \varphi)^* \varphi\Bigr) = 0 \; . \label{Aeqn}
\end{eqnarray}
Here the covariant derivative operator is $D_{\mu} \!\equiv\! \partial_{\mu}
\!+\! i e A_{\mu}$. The result of integrating (\ref{phieqn}-\ref{Aeqn}) is
the Yang-Feldman equations \cite{YF},
\begin{eqnarray}
\varphi(x) & = & \varphi_0(x) + \int_{\eta_i}^0 d\eta' \int d^{D-1}x' G(x;x') 
I[\varphi^*,\varphi,A](x') \; , \label{phi0} \\
A_{\mu}(x) & = & A_{0\mu}(x) + \int_{\eta_i}^0 \int d^{D-1}x' \Bigl[\mbox{}_{
\mu} G_{\nu} \Bigr](x;x') J^{\nu}[\varphi^*,\varphi,A](x') \; . \label{A0}
\end{eqnarray}
The two interactions are,
\begin{eqnarray}
I[\varphi^*,\varphi,A] & = & -ie A_{\mu} \sqrt{-g} g^{\mu\nu} \partial_{\nu}
\varphi \!-\! ie \partial_{\mu} (\sqrt{-g} g^{\mu\nu} A_{\nu} \varphi) 
\nonumber \\
& & \hspace{.5cm} + \sqrt{-g} g^{\mu\nu} e^2 A_{\mu} A_{\nu} \varphi
+ \frac{\delta \xi \sqrt{-g} R \varphi}{1 \!+\! \delta Z_2} + \frac{\delta
\lambda \sqrt{-g} \varphi^* \varphi^2}{2 (1 \!+\! \delta Z_2)} \; , \qquad \\
J^{\mu}[\varphi^*,\varphi,A] & = & \frac{-i e \sqrt{-g} g^{\mu\nu}}{1 \!+\!
\delta Z_3} \Bigl(\varphi^* D_{\nu} \varphi \!-\! (D_{\nu} \varphi)^*
\varphi\Bigr) \; . 
\end{eqnarray}
$G(x;x')$ is any solution to the free scalar Green's function equation,
\begin{equation}
\partial_{\mu} \Bigl(\sqrt{-g} g^{\mu\nu} \partial_{\nu} G(x;x') \Bigr) = 
\delta^D(x \!-\! x') \; ,
\end{equation}
and $[\mbox{}_{\mu} G_{\nu}](x;x')$ is any solution to the analogous photon 
Green's function equation in whatever gauge is employed. For flat space 
scattering problems the Green's functions would obey Feynman boundary
conditions --- and hence amount to $-i$ times the Feynman propagators.
In cosmology it is more natural to use retarded boundary conditions.

We stress that the fundamental operators $\varphi(x)$ and $A_{\mu}(x)$ are 
unique and unaffected by the choices of $\eta_i$ and the boundary conditions
for the Greens functions. What changes as we vary $\eta_i$ and the Green's 
functions is the free fields, $\varphi_0(x)$ and $A_{0\mu}(x)$. The fact that 
they obey the linearized equations of motion and agree with the full fields 
at $\eta \!=\! \eta_i$ implies that they can be expanded in terms of free 
creation and annihilation operators,
\begin{eqnarray}
\varphi_0(x) & \!\!\!\!\!=\!\!\!\!\! & \int \!\! \frac{d^{D-1}k}{(2\pi)^{D-1}} 
\Bigl\{u(\eta,k) e^{i \vec{k} \cdot \vec{x}} \alpha(\vec{k}) + u^*(\eta,k) 
e^{-i\vec{k} \cdot \vec{x}} \beta^{\dagger}(\vec{k}) \Bigr\} \; , \\
A_{0\mu}(x) & \!\!\!\!\!=\!\!\!\!\! & \int \!\!\frac{d^{D-1}k}{(2\pi)^{D-1}} 
\!\! \sum_{\lambda} \Bigl\{\epsilon_{\mu}(\eta,k,\lambda) e^{i \vec{k} \cdot 
\vec{x}} \gamma(\vec{k},\lambda) + \epsilon_{\mu}^*(\eta,k,\lambda) 
e^{-i\vec{k} \cdot \vec{x}} \gamma^{\dagger}(\vec{k},\lambda) \Bigr\} \; . 
\qquad
\end{eqnarray}
Here $u(\eta,k)$ and $\epsilon_{\mu}(\eta,k,\lambda)$ are the Bunch-Davies
mode functions \cite{BD} in whatever gauge is being used. They also have 
canonically normalized Wronskians, which for the scalar is,
\begin{equation}
u(\eta,k) \partial_0 u^*(\eta,k) - \partial_0 u(\eta,k) u^*(\eta,k) = 
\frac{i}{a^{D-2}} \; . 
\end{equation}
Although the various creation and annihilation operators change as different 
Green's functions are employed in (\ref{phi0}) and (\ref{A0}), their nonzero 
commutation relations remain fixed,
\begin{eqnarray}
[\alpha(\vec{k}),\alpha^{\dagger}(\vec{k}')] & = & (2\pi)^{D-1}
\delta^{D-1}(\vec{k} \!-\! \vec{k}') = [\beta(\vec{k}),\beta^{\dagger}(
\vec{k}')] \; , \qquad \\
\Bigl[\gamma(\vec{k},\lambda),\gamma^{\dagger}(\vec{k}',\lambda')\Bigr] & = &
(2\pi)^{D-1} \delta^{D-1}(\vec{k} \!-\! \vec{k}') \delta_{\lambda\lambda'} \; .
\end{eqnarray}

One develops perturbation theory by iterating (\ref{phi0}-\ref{A0}) to 
obtain expansions for the full fields, $\varphi(x)$ and $A_{\mu}(x)$, in 
terms of the free fields, $\varphi_0(x)$ and $A_{0\mu}(x)$. With our choice
of retarded boundary conditions the expansion for $\varphi(x)$ takes the form,
\begin{eqnarray}
\varphi(x) & = & \varphi_0(x) + \int_{\eta_i}^0 d\eta' \int d^{D-1}x' 
G_{\rm ret}(x;x') I[\varphi^*,\varphi,A](x') \; , \\
& = & \varphi_0(x) + \int_{\eta_i}^0 d\eta' \int d^{D-1}x' 
G_{\rm ret}(x;x') I[\varphi^*_0,\varphi_0,A_0](x') + \ldots \; .
\end{eqnarray}
With our choice of $\vert \Psi_f\rangle \!=\! \vert \Psi_i\rangle$ as free
vacuum at $\eta_i$, the quantum corrected scalar mode function is,
\begin{equation}
\Phi(x;\vec{k}) = \Bigl\langle \Omega \Bigl\vert \Bigl[\varphi(x),
\alpha^{\dagger}(\vec{k})\Bigr] \Bigr\vert \Omega \Bigr\rangle =
u(\eta,k) e^{i\vec{k} \cdot \vec{x}} + O(e^2) \; . \label{effmod}
\end{equation}

\subsection{The Schwinger-Keldysh Formalism}

It is straightforward to demonstrate \cite{MW3} that the quantum corrected 
mode function (\ref{effmod}) obeys the linearized effective field equations 
of the Schwinger-Keldysh formalism \cite{JS,KTM,BM,LVK}. This is a covariant
extension of Feynman diagrams which produces true expectation values rather 
than the in-out matrix elements of conventional Feynman diagrams. Because 
many excellent reviews of this subject exist \cite{CSHY,RDJ,CH,FW,SW1} we
will simply state that part of the formalism which is necessary for this
work. 

The basic idea is that the endpoints of propagators acquire a $\pm$
polarity, so every propagator $i\Delta(x;x')$  of the in-out formalism 
generalizes to four Schwinger-Keldysh propagators: $i\Delta_{\scriptscriptstyle
++}(x;x')$, $i\Delta_{\scriptscriptstyle +-}(x;x')$, $i\Delta_{
\scriptscriptstyle -+}(x;x')$ and $i\Delta_{\scriptscriptstyle --}(x;x')$.
The usual vertices attach to $+$ endpoints, whereas $-$ endpoints attach to
the conjugate vertices. Had the state been some perturbative correction of 
free vacuum, this correction would give rise to additional vertices on the
initial value surface --- $+$ ones from $\Psi_i[\varphi^*_+(\eta_i),
\varphi_+(\eta_i)]$ and $-$ ones from $\Psi_f[\varphi_-^*(\eta_i),
\varphi_-(\eta_i)]$.

Because each external line can be either $+$ or $-$ in the Schwinger-Keldysh
formalism, each $N$-point 1PI function of the in-out formalism corresponds to 
$2^N$ Schwinger-Keldysh $N$-point functions. The Schwinger-Keldysh effective 
action is the generating functional of these 1PI functions, so it depends upon 
$+$ fields and $-$ fields. The effective field equations come from varying 
with respect to either polarity and then setting the two polarities equal,
\begin{eqnarray}
\lefteqn{\frac{\delta \Gamma[\varphi_{\scriptscriptstyle \pm}^*,\varphi_{
\scriptscriptstyle\pm}]}{\delta \varphi^*_{\scriptscriptstyle +}(x)}
\Biggl\vert_{\varphi_{\scriptscriptstyle \pm} = \varphi \atop \varphi_{
\scriptscriptstyle \pm}^* = \varphi^*} = \partial_{\mu} \Bigl(\sqrt{-g} 
g^{\mu\nu} \partial_{\nu} \varphi(x)\Bigr) } \nonumber \\
& & \hspace{1.4cm} - \int_{\eta_i}^0 d\eta' \int d^3x' \Bigl\{M^2_{
\scriptscriptstyle ++}(x;x') + M^2_{\scriptscriptstyle +-}(x;x')\Bigr\} 
\varphi(x') + O\Bigl(\varphi^* \varphi^2 \Bigr) \; . \qquad
\end{eqnarray}
Note that we have taken the regularization parameter $D$ to its unregulated
value of $D \!=\! 4$, in view of the fact that the self-mass-squared is
assumed to be fully renormalized. It is the {\it linearized} effective field 
equation which $\Phi(x;\vec{k})$ obeys,
\begin{equation}
\partial_{\mu} \Bigl(\sqrt{-g} g^{\mu\nu} \partial_{\nu} \Phi(x;\vec{k})\Bigr) 
- \int_{\eta_i}^0 d\eta' \int d^3x' \Bigl\{M_{\scriptscriptstyle ++}(x;x') 
+ M_{\scriptscriptstyle +-}(x;x') \Bigr\} \Phi(x';\vec{k}) = 0 \; . 
\label{lineqn}
\end{equation}

Converting the single, in-out self-mass-squared --- $M^2(x;x')$
--- into the four polarities of the Schwinger-Keldysh formalism is simple. The
$++$ polarization is the usual in-out self-mass-squared and the $--$
polarization is minus its conjugate,
\begin{equation}
M^2_{\scriptscriptstyle ++}(x;x') = M^2(x;x') \qquad {\rm and} \qquad
M^2_{\scriptscriptstyle --}(x;x') = - \Bigl(M^2(x;x')\Bigr)^* \; .
\end{equation}
To obtain the mixed polarizations we observe that, in de Sitter conformal
coordinates, the in-out result depends mostly upon the conformal coordinate 
interval \cite{MW3},
\begin{equation}
\Delta x^2_{\scriptscriptstyle ++}(x;x') \equiv \Bigl\Vert \vec{x} \!-\! 
\vec{x}' \Bigr\Vert^2 - \Bigl(\vert \eta \!-\! \eta'\vert \!-\! i \delta
\Bigr)^2 \; .
\end{equation}
To obtain the $+-$ and $-+$ polarities at one loop order we must do three 
things \cite{MW3}:
\begin{itemize}
\item{Replace $\Delta x^2_{\scriptscriptstyle ++}(x;x')$ with the appropriate
coordinate interval,
\begin{eqnarray}
\Delta x^2_{\scriptscriptstyle +-}(x;x') & \equiv & \Bigl\Vert \vec{x} \!-\! 
\vec{x}' \Bigr\Vert^2 - \Bigl(\eta \!-\! \eta' \!+\! i \delta \Bigr)^2 \; , \\
\Delta x^2_{\scriptscriptstyle -+}(x;x') & \equiv & \Bigl\Vert \vec{x} \!-\! 
\vec{x}' \Bigr\Vert^2 - \Bigl(\eta \!-\! \eta' \!-\! i \delta \Bigr)^2 \; ;
\end{eqnarray}}
\item{Drop all delta function terms; and}
\item{Multiply the result by $-1$ to account for the $-$ vertex.}
\end{itemize}
It will be seen that the $++$ and $+-$ terms in (\ref{lineqn}) exactly
cancel for $\eta' \!>\! \eta$ and also, in the limit $\delta \!\rightarrow\! 
0$, for $x^{\prime \mu}$ outside the light-cone of $x^{\mu}$. This is how the
Schwinger-Keldysh formalism gives causal effective field equations.

\subsection{The Meaning of ``Solve''}

Two important limitations on our current knowledge restrict the sense in
which we can usefully solve the effective mode equation. The first of 
these is that we do not know the exact scalar self-mass-squared. The full
result can be expressed as a series in powers of the loop counting parameter 
$e^2$,
\begin{equation}
M_{\scriptscriptstyle ++}(x;x') + M_{\scriptscriptstyle +-}(x;x') =
\sum_{\ell=1}^{\infty} e^{2\ell} \mathcal{M}^2_{\ell}(x;x') \; .
\end{equation}
Because we possess only the $\ell \!=\! 1$ term, we can only solve the
effective mode equation (\ref{lineqn}) perturbatively to order $e^2$. 
That is, we substitute a series solution,
\begin{equation}
\Phi(x;\vec{k}) \equiv u(\eta,k) e^{i\vec{k} \cdot \vec{x}} + \sum_{\ell=1}^{
\infty} e^{2\ell} \Phi_{\ell}(\eta,k) e^{i \vec{k} \cdot\vec{x}} \; , 
\end{equation}
and then segregate according to powers of $e^2$. The 0th order solution is
well known,
\begin{equation}
u(\eta,k) = \frac{H}{\sqrt{2 k^3}} \Bigl[1 \!-\! \frac{ik}{H a}\Bigr]
\exp\Bigl[\frac{ik}{H a}\Bigr] \; , . \label{uform}
\end{equation}
The 1st order solution $\Phi_1(\eta,k)$ obeys the equation,
\begin{equation}
a^2 \Bigl[\partial_0^2 \!+\! 2 H a \partial_0 \!+\! k^2\Bigr] \Phi_1(\eta,k)
= -\int_{\eta_i}^0 d\eta' \int d^3x' \, \mathcal{M}_1^2(x;x') u(\eta',k)
e^{i\vec{k} \cdot (\vec{x} - \vec{x}')} \; . \label{oureqn}
\end{equation}

The second limitation is indicated by the subscript ``$\eta_i$'' on the
temporal integration in (\ref{oureqn}): we release the universe in free 
vacuum at time $\eta \!=\! \eta_i$. Not much is known about the wave 
functionals of interacting quantum field theories in curved space, but it
can hardly be that free vacuum is very realistic. All the finite energy
states of interacting flat space quantum field theories possess important 
corrections. It is inconceivable that similar corrections are not present
in curved space, at least in the far ultraviolet regime for which the
geometry is effectively flat. 

Although one could perturbatively correct the free states to a few orders, 
just as in nonrelativistic quantum mechanics, the standard procedure of flat 
space quantum field theory is to instead release the system in free vacuum 
at asymptotically early times. In the weak operator sense, infinite time 
evolution resolves the difference between free vacuum and true vacuum into 
shifts of the mass, field strength and background field \cite{BjD}. In 
cosmology we cannot typically employ this procedure, however, it is still
possible to perturbatively correct the state wave functionals. 

Corrections to the initial state would show up as new interaction vertices 
on the initial value surface. One would expect them to have a large effect on
the expectation values of operators near the initial value. One would also
expect their effects to decay in the expectation values of late time 
operators. We have not worked out these surface vertices but the need for
them has been apparent from the very first fully regulated computations of
this type \cite{OW1,PTW2}. For example, consider the expectation value of 
the stress energy tensor of a massless, minimally coupled scalar with a
quartic self interaction on a locally de Sitter background. If the state is 
released in free vacuum at the instant when the de Sitter scale factor takes
the value $a \!=\! 1$, then one can choose the renormalization parameters
so that the energy density and pressure induced at two loop order are
\cite{OW2},
\begin{eqnarray}
\rho & = & \frac{\lambda H^4}{(2 \pi)^4} \Biggl\{\frac18 \ln^2(a) \!+\! 
\frac{a^{-3}}{18} \!-\! \frac18 \sum_{n=1}^{\infty} \frac{(n\!+\!2) a^{-n-1}}{
(n\!+\!1)^2} \Biggr\} + O(\lambda^2) \; , \\
p & = & \frac{\lambda H^4}{(2 \pi)^4} \Biggl\{-\frac18 \ln^2(a) \!-\! 
\frac1{12} \ln(a) \!-\! \frac1{24} \sum_{n=1}^{\infty} \frac{(n^2\!-\!4) 
a^{-n-1}}{(n\!+\!1)^2} \Biggr\} + O(\lambda^2) \; .
\end{eqnarray}
It is the terms which fall off like powers of $1/a$ that we suspect can be
absorbed into an order $\lambda$ correction of the initial state,
\begin{eqnarray}
\Delta \rho & = & \frac{\lambda H^4}{(2 \pi)^4} \Biggl\{ \frac{a^{-3}}{18} 
\!-\! \frac18 \sum_{n=1}^{\infty} \frac{(n\!+\!2) a^{-n-1}}{(n\!+\!1)^2} 
\Biggr\} + O(\lambda^2) \; , \label{Drho} \\
\Delta p & = & \frac{\lambda H^4}{(2 \pi)^4} \Biggl\{- \frac1{24} 
\sum_{n=1}^{\infty} \frac{(n^2\!-\!4) a^{-n-1}}{(n\!+\!1)^2} \Biggr\} 
+ O(\lambda^2) \; . \label{Dp}
\end{eqnarray}
Of course the fact that they fall off as one evolves away from the initial
value surface suggests that they can be absorbed into some kind of local
interaction there. Note also that they are separately conserved, which is
exactly what would be the case if they could be canceled by a new interaction 
vertex.

Note particularly that expressions (\ref{Drho}-\ref{Dp}) {\it diverge}
on the initial value surface. Divergences on the initial value surface have
also be found in the effective mode equations for photons in SQED 
\cite{PTW2,PW2}, for fermions in Yukawa theory \cite{PW1}, for scalars in 
Yukawa theory \cite{DW}, and for fermions in Dirac + Einstein \cite{MW3}. 
These initial value divergences reflect the fact that free vacuum is very
far away from any physically accessible state. Equation (\ref{oureqn}) 
accurately determines the one loop correction to the mode function
(\ref{effmod}) appropriate to free vacuum, however, that mode function
has little physical relevance because free vacuum can not be assembled.

If one desires expressions for physically relevant mode functions which are
valid even for times near the initial value surface, there is no alternative 
to including corrections to the state wave functional in the self-mass-squared.
This seems quite practicable because one need only do it perturbatively to the
same order as the ordinary, ``volume'' contributions are known. Although there
are no stationary states for SQED on de Sitter background, it would probably 
be an excellent approximation --- becoming exact in the ultraviolet --- to 
simply solve for the relevant corrections in flat space. 

Unfortunately, we do not now possess the order $\lambda$ correction to the
state wave functional. That means there is absolutely no point in trying to 
solve (\ref{oureqn}) for all times. However, it does not mean (\ref{oureqn})
is devoid of physical information. In particular, the effects of the state
corrections must fall off at late times --- and quite rapidly, like inverse 
powers of the scale factor. This fall-off is evident in expressions 
(\ref{Drho}-\ref{Dp}). It simply reflects the same process which is employed
in flat space quantum field theory \cite{BjD} whereby time evolution 
washes away the difference between free vacuum and true vacuum. We can 
therefore extract valid information from (\ref{oureqn}) by solving it in the
late time limit.

\section{Non-de Sitter Invariant Gauge}

Reliable results for the one loop scalar self-mass-squared have been
obtained in two gauges. The first is a non-de Sitter invariant analogue 
of Feynman gauge which corresponds to adding the gauge-fixing term \cite{KW},
\begin{equation}
\mathcal{L}_{\rm GF} = -\frac12 a^{D-4} \Bigl(\eta^{\mu\nu} A_{\mu , \nu}
\Bigr)^2 \; .
\end{equation}
In this gauge the renormalized in-out (and hence also, the $++$)
self-mass-squared is,
\begin{eqnarray}
\lefteqn{M^2_{\mbox{\tiny non} \atop \scriptscriptstyle ++}(x;x') = - e^2 
\delta Z^{\rm non}_{\rm fin} \sqrt{-g} \square \delta^4(x \!-\! x') 
+ 12 e^2 H^2 \delta \xi^{\rm inv}_{\rm fin} \sqrt{-g} \delta^4(x \!-\! x') } 
\nonumber \\
& & + \frac{e^2 a a' }{8 \pi^2} \, \ln(a a') \partial^2 \delta^4(x\!-\!x') 
- \frac{e^2 H^2}{4 \pi^2} \, a^4 \ln(a) \delta^4(x\!-\!x') \nonumber \\
& & \hspace{1cm} - \frac{i e^2 a a'}{2^8 \pi^4} \, \partial^6 \Biggl\{ 
\ln^2\Bigl(\frac14 H^2 \Delta x^2_{\scriptscriptstyle ++}\Bigr) \!-\! 2 
\ln\Bigl(\frac14 H^2 \Delta x^2_{\scriptscriptstyle ++}\Bigr) \Biggr\} 
\nonumber \\
& & \hspace{2cm} - \frac{i e^2 H^4}{2^6 \pi^4} \, (a a')^3 \Biggl\{ \nabla^2 
\Bigl[ \ln^2\Bigl(\frac14 H^2 \Delta x^2_{\scriptscriptstyle ++}\Bigr) 
\!-\! \ln\Bigl(\frac14 H^2 \Delta x^2_{\scriptscriptstyle ++} \Bigr) \Bigr] 
\nonumber \\
& & \hspace{3cm} + \frac{\partial^2}2 \Bigl[\ln^2\Bigl(\frac14 H^2\Delta 
x^2_{\scriptscriptstyle ++}\Bigr) \!-\! 3 \ln\Bigl(\frac14 H^2 \Delta 
x^2_{\scriptscriptstyle ++}\Bigr) \Bigr] \Biggr\} + O(e^4) \; . \qquad 
\label{non++}
\end{eqnarray}
Here $e^2 \delta Z^{\rm non}_{\rm fin}$ and $e^2 \delta \xi^{\rm non}_{\rm 
fin}$ are the arbitrary finite parts of $\delta Z_2$ and $\delta \xi$, in this
gauge and at order $e^2$. The various derivative operators are,
\begin{equation}
\square \equiv \frac1{\sqrt{-g}} \partial_{\mu} \Bigl(\sqrt{-g} g^{\mu\nu}
\partial_{\nu}\Bigr) \quad , \quad \partial^2 \equiv \eta^{\mu\nu} 
\partial_{\mu} \partial_{\nu} \quad {\rm and} \quad \nabla^2 \equiv 
\partial_i \partial_i \; .
\end{equation}
Applying the rules of the previous section allows us to recognize the $+-$
self-mass-squared as,
\begin{eqnarray}
\lefteqn{M^2_{\mbox{\tiny non} \atop \scriptscriptstyle +-}(x;x') = 
\frac{i e^2 a a'}{2^8 \pi^4} \, \partial^6 \Biggl\{ \ln^2\Bigl(\frac14 H^2 
\Delta x^2_{\scriptscriptstyle +-}\Bigr) \!-\! 2 \ln\Bigl(\frac14 H^2 
\Delta x^2_{\scriptscriptstyle +-}\Bigr) \Biggr\} } \nonumber \\
& & \hspace{2cm} + \frac{i e^2 H^4}{2^6 \pi^4} \, (a a')^3 \Biggl\{ \nabla^2 
\Bigl[ \ln^2\Bigl(\frac14 H^2 \Delta x^2_{\scriptscriptstyle +-}\Bigr) 
\!-\! \ln\Bigl(\frac14 H^2 \Delta x^2_{\scriptscriptstyle +-} \Bigr) \Bigr] 
\nonumber \\
& & \hspace{3cm} + \frac{\partial^2}2 \Bigl[\ln^2\Bigl(\frac14 H^2\Delta 
x^2_{\scriptscriptstyle +-}\Bigr) \!-\! 3 \ln\Bigl(\frac14 H^2 \Delta 
x^2_{\scriptscriptstyle +-}\Bigr) \Bigr] \Biggr\} + O(e^4) \; . \qquad 
\label{non+-}
\end{eqnarray}

The next step is to combine (\ref{non++}) with (\ref{non+-}) to read off 
$\mathcal{M}^2_1(x;x')$ for the right hand side of (\ref{oureqn}). To 
simplify the notation we first define the spatial and temporal intervals,
\begin{equation}
\Delta r \equiv \Vert \vec{x} \!-\! \vec{x}'\Vert \qquad {\rm and} \qquad
\Delta \eta \equiv \eta \!-\! \eta' \; .
\end{equation}
We also take note of the differences of powers of $++$ and $+-$ logarithms,
\begin{eqnarray}
\ln\Bigl(\frac14 H^2 \Delta x^2_{\scriptscriptstyle ++}\Bigr) \!-\!
\ln\Bigl(\frac14 H^2 \Delta x^2_{\scriptscriptstyle +-}\Bigr) & \!\!\!=\!\!\! &
2\pi i \theta(\Delta \eta \!-\! \Delta r) \; , \label{theta1} \\
\ln^2\Bigl(\frac14 H^2 \Delta x^2_{\scriptscriptstyle ++}\Bigr) \!-\!
\ln^2\Bigl(\frac14 H^2 \Delta x^2_{\scriptscriptstyle +-}\Bigr) &\!\!\!=\!\!\!&
4\pi i \theta(\Delta \eta \!-\! \Delta r) \ln\Bigl[\frac{H^2}4 (\Delta \eta^2
\!-\! \Delta r^2)\Bigr] . \qquad \label{theta2}
\end{eqnarray}
With this notation we find,
\begin{eqnarray}
\lefteqn{\mathcal{M}^2_{{\rm non}1}(x;x') = - \delta Z^{\rm non}_{\rm fin} 
\sqrt{-g} \square \delta^4(x \!-\! x') + 12 H^2 \delta \xi^{\rm non}_{\rm fin} 
\sqrt{-g} \delta^4(x \!-\! x') } \nonumber \\
& & + \frac{a a'}{8 \pi^2} \, \ln(a a') \partial^2 \delta^4(x\!-\!x') 
- \frac{H^2}{4 \pi^2} \, a^4 \ln(a) \delta^4(x\!-\!x') \nonumber \\
& & \hspace{1cm} + \frac{a a'}{2^6 \pi^3} \, \partial^6 \Biggl\{ 
\theta(\Delta \eta \!-\! \Delta r) \Biggl(\ln\Bigl[\frac{H^2}4 (\Delta \eta^2
\!-\! \Delta r^2)\Bigr] \!-\! 1\Biggr) 
\Biggr\} \nonumber \\
& & \hspace{2cm} + \frac{H^4}{2^4 \pi^3} \, (a a')^3 \Biggl\{ \nabla^2 
\Biggl[ \theta(\Delta \eta \!-\! \Delta r) \Biggl(\ln\Bigl[\frac{H^2}4 (\Delta 
\eta^2 \!-\! \Delta r^2)\Bigr] \!-\! \frac12\Biggr) \Biggr] \nonumber \\
& & \hspace{3.5cm} + \frac{\partial^2}2 \Biggl[ \theta(\Delta \eta \!-\! 
\Delta r) \Biggl(\ln\Bigl[\frac{H^2}4 (\Delta \eta^2 \!-\! \Delta r^2)\Bigr] 
\!-\! \frac32\Biggr) \Biggr] \Biggr\} . \qquad \label{calM}
\end{eqnarray}
Note that the result is manifestly real and that each of the nonlocal factors 
is shielded by a causality-preserving factor of $\theta(\Delta \eta \!-\! 
\Delta r)$. These are features of the Schwinger-Keldysh formalism which are
absent from the usual (in-out) effective field equations.

We now evaluate the contribution each term in (\ref{calM}) makes to the
right hand side of the one loop effective mode equation (\ref{oureqn}). The
local contributions are simple,
\begin{eqnarray}
\lefteqn{-\int_{\eta_i}^{0} \!\! d\eta' \!\! \int \!\! d^3x' \Biggl\{-\delta
Z^{\rm non}_{\rm fin} a^4 \square \delta^4(x \!-\! x') \, u(\eta',k) 
e^{i \vec{k} \cdot (\vec{x}' - \vec{x})} \Biggr\} } \nonumber \\
& & \hspace{1cm} = \delta Z^{\rm non}_{\rm fin} a^4 \square \Bigl[u(\eta,k) 
e^{i \vec{k} \cdot\vec{x}} \Bigr] e^{-i \vec{k} \cdot \vec{x}} = 0 \; , 
\qquad \\
\lefteqn{-\int_{\eta_i}^{0} \!\! d\eta' \!\! \int \!\! d^3x' \Biggl\{12 H^2 
\delta \xi^{\rm non}_{\rm fin} a^4 \delta^4(x \!-\! x') \, u(\eta',k) e^{i i
\vec{k} \cdot (\vec{x}' - \vec{x})} \Biggr\} } \nonumber \\
& & \hspace{1cm} = \frac{H^2 a^4}{4 \pi^2} \Bigl\{- 48 \pi^2 \delta \xi^{\rm
non}_{\rm fin} u(\eta,k) \Bigr\} \; , \qquad \\
\lefteqn{-\int_{\eta_i}^{0} \!\! d\eta' \!\! \int \!\! d^3x' \Biggl\{ 
\frac{a a'}{8 \pi^2} \ln(a a') \partial^2 \delta^4(x \!-\! x') \, u(\eta',k) 
e^{i \vec{k} \cdot (\vec{x}' - \vec{x})} \Biggr\} } \nonumber \\
& & \hspace{1cm} = \frac{a}{8\pi^2} \Biggl\{\ln(a) \Bigl(\partial_0^2 \!+\! 
k^2\Bigr) \Bigl(a u(\eta,k)\Bigr) \!+\! \Bigl(\partial_0^2 \!+\! k^2\Bigr) 
\Bigl( a \ln(a) u(\eta,k)\Bigr) \Biggr\} \; , \qquad \\
& & \hspace{1cm} = \frac{H^2 a^4}{4 \pi^2} \Biggl\{ 2 \ln(a) u(\eta,k) \!+\! 
\frac32 u(\eta,k) + \frac1{H a} \partial_0 u(\eta,k) \Biggr\} \; , \qquad \\
\lefteqn{-\int_{\eta_i}^{0} \!\! d\eta' \!\! \int \!\! d^3x' \Biggl\{
-\frac{H^2}{4 \pi^2} a^4 \ln(a) \delta^4(x \!-\! x') \, u(\eta',k) e^{i \vec{k} 
\cdot (\vec{x}' - \vec{x})} \Biggr\} } \nonumber \\
& & \hspace{1cm} = \frac{H^2 a^4}{4\pi^2} \Bigl\{ \ln(a) u(\eta,k) \Bigr\}
\; . \qquad
\end{eqnarray}

The contribution of the first nonlocal term in (\ref{calM}) has been evaluated 
in a previous study of one loop corrections to the scalar mode functions of 
Yukawa theory \cite{DW}. By making the replacements,
\begin{equation}
f \longrightarrow 1 \qquad , \qquad \mu \longrightarrow \frac{H}2 \qquad 
{\rm and} \qquad g(\eta',k) \longrightarrow u(\eta',k) \; ,
\end{equation}
we can read off the result from equation (68) of that paper, and some of the 
subsequent asymptotic expansions,
\begin{eqnarray}
\lefteqn{-\frac{a}{2^6 \pi^3} e^{-i\vec{k} \cdot \vec{x}} \partial^6 \!\!
\int_{\eta_i}^{\eta} \!\! d\eta' a' u(\eta',k) \int_{\Delta r \leq \Delta \eta} 
\!\!\!\!\!\!\!\!\!\!\!\! d^3x' \, e^{i\vec{k} \cdot \vec{x}'} \Biggl\{ 
\ln\Bigl[\frac14 H^2 (\Delta \eta^2 \!-\! \Delta r^2)\Bigr] \!-\! 1\Biggr\} } 
\nonumber \\
& & \hspace{1cm} = \frac{a}{4 \pi^2} (\partial_0^2 \!+\! k^2) \left\{ \matrix{
-a \ln(a) u(\eta,k) \cr + \partial_0 \int_{\eta_i}^{\eta} d\eta' a' u(\eta',k) 
\cos(k \Delta \eta) \ln(1 \!-\! \frac{a'}{a})\cr+ k \int_{\eta_i}^{\eta} d\eta'
a' u(\eta',k) \sin(k \Delta \eta) \ln(1 \!-\! \frac{a'}{a}) } \right\} \; , \\
& & \hspace{1cm} = \frac{H^2 a^4}{4 \pi^2} \Biggl\{-2 \ln(a) u(\eta,k) \!-\! 
3 u(\eta,k) \!+\! O\Bigl(\frac1{a^2}\Bigr) \Biggr\} . \qquad
\end{eqnarray}

We evaluate the contribution of the second nonlocal term in (\ref{calM})
by first making the change of variable $\vec{r} \!=\! \vec{x}' \!-\! \vec{x}$,
then performing the angular integrations, and finally changing the radial 
variable to $r \!=\! \Delta \eta z$,
\begin{eqnarray}
\lefteqn{-\frac{H^4 a^3}{2^4 \pi^3} e^{-i\vec{k} \cdot \vec{x}} \nabla^2 \!\!
\int_{\eta_i}^{\eta} \!\! d\eta' a^{\prime 3} u(\eta',k) \int_{\Delta r \leq 
\Delta \eta} \!\!\!\!\!\!\!\!\!\!\!\!\! d^3x' \, e^{i\vec{k} \cdot \vec{x}'} 
\Biggl\{ \ln\Bigl[\frac14 H^2 (\Delta \eta^2 \!-\! \Delta r^2)\Bigr] \!-\! 
\frac12 \Biggr\} } \nonumber \\
& & \hspace{-.7cm} = \!\frac{H^4 a^3}{4 \pi^2} k \!\! \int_{\eta_i}^{\eta} 
\!\!\!\! d\eta' a^{\prime 3} u(\eta',k) \Delta \eta^2 \!\!\! \int_0^1 \!\!\!\! 
dz \, z \sin(k \Delta \eta z) \! \Biggl\{\!2 \ln(H \Delta \eta) \!+\! 
\ln\Bigl(\!\frac{1 \!-\!  z^2}4 \!\Bigr) \!-\! \frac12 \!\Biggr\} . \quad
\end{eqnarray}
The $z$ integration can be performed in terms of sine and cosine integrals
but there is no need to do this. We simply observe the behavior of the
$z$ integral for small $\Delta \eta$,
\begin{equation}
\int_0^1 \!\!\! dz \, z \sin(k \Delta \eta z) \Biggl\{2 \ln(H \Delta \eta) 
\!+\! \ln\Bigl(\!\frac{1 \!-\!  z^2}4 \!\Bigr) \!-\! \frac12 \Biggr\} 
\longrightarrow \frac23 k \Delta \eta \ln(H \Delta \eta) \; . \label{intform}
\end{equation}
This implies that the $\eta'$ integration in (\ref{intform}) converges at 
arbitrarily late times ($\eta \!\rightarrow \! 0^-$). Hence the second 
nonlocal term can be expanded as follows,
\begin{eqnarray}
\lefteqn{-\frac{H^4 a^3}{2^4 \pi^3} e^{-i\vec{k} \cdot \vec{x}} \nabla^2 \!\!
\int_{\eta_i}^{\eta} \!\! d\eta' a^{\prime 3} u(\eta',k) \int_{\Delta r \leq 
\Delta \eta} \!\!\!\!\!\!\!\!\!\!\!\!\! d^3x' \, e^{i\vec{k} \cdot \vec{x}'} 
\Biggl\{ \ln\Bigl[\frac14 H^2 (\Delta \eta^2 \!-\! \Delta r^2)\Bigr] \!-\! 
\frac12 \Biggr\} } \nonumber \\
& & \hspace{7cm} = \frac{H^2 a^3}{4 \pi^2} \Biggl\{ K + 
O\Bigl(\frac{\ln(a)}{a} \Bigr) \Biggr\} , \qquad 
\end{eqnarray}
where the constant $K$ is,
\begin{equation}
K = \frac{k}{H^2} \int_{a_i}^{\infty} \!\!\! da' \frac{u(\eta',k)}{a'} 
\!\! \int_0^1 \!\! dz \, z \sin\Bigl(\frac{k z}{H a'}\Bigr) \Biggl\{-2 \ln(a')
\!+\! \ln\Bigl(\frac{1 \!-\! z^2}4\Bigr) \!-\! \frac12\Biggr\} .
\end{equation}
We will see that only terms which grow as fast as $a^4$ can give significant
corrections at late times.

The contribution from the final nonlocal term in (\ref{calM}) is the most 
difficult to evaluate. The initial steps are the same as for the second term.
However, one must then act the temporal derivatives and express the $z$ 
integration in terms of sine and cosine integrals,
\begin{eqnarray}
\lefteqn{-\frac{H^4 a^3}{2^5 \pi^3} e^{-i\vec{k} \cdot \vec{x}} \partial^2 \!\!
\int_{\eta_i}^{\eta} \!\! d\eta' a^{\prime 3} u(\eta',k) \int_{\Delta r \leq 
\Delta \eta} \!\!\!\!\!\!\!\!\!\!\!\!\! d^3x' \, e^{i\vec{k} \cdot \vec{x}'} 
\Biggl\{ \ln\Bigl[\frac14 H^2 (\Delta \eta^2 \!-\! \Delta r^2)\Bigr] \!-\! 
\frac32 \Biggr\} } \nonumber \\
& & = \frac{H^4 a^3}{8 \pi^2 k} (\partial_0^2 \!+\! k^2) \!\! 
\int_{\eta_i}^{\eta} \!\!\!\! d\eta' a^{\prime 3} u(\eta',k) \Delta \eta^2 
\nonumber \\
& & \hspace{3cm} \times \int_0^1 \!\!\! dz \, z \sin(k \Delta \eta z) 
\Biggl\{2 \ln(H \Delta \eta) \!+\! \ln\Bigl(\frac{1 \!-\!  z^2}4 \Bigr) \!-\! 
\frac32 \!\Biggr\} , \qquad \\
& & = \frac{H^4 a^3}{8 \pi^2 k} \!\! \int_{\eta_i}^{\eta} \!\!\!\! d\eta' 
a^{\prime 3} u(\eta',k) \Biggl\{-2 \cos(\alpha) \int_0^{2\alpha} \!\!\! dt \, 
\frac{\sin(t)}{t} \nonumber \\
& & \hspace{4cm} + 2 \sin(\alpha) \Biggl[ \int_0^{2 \alpha} \!\!\! dt \,
\frac{\cos(t) \!-\! 1}{t} \!+\! 2 \ln\Bigl(\frac{H \alpha}{k}\Bigr) \!-\!
\frac12\Biggr] \Biggr\} . \qquad \label{exact}
\end{eqnarray}
Here we define $\alpha \equiv k \Delta \eta$.

To extract the leading late time behavior of expression (\ref{exact}) we 
first note that the term in curly brackets vanishes like $4 \alpha 
\ln(\alpha)$ for $\eta' \!=\! \eta$. This means that the upper limit makes
no contribution if we integrate by parts on $a^{\prime 3}$,
\begin{eqnarray}
\lefteqn{\int_{\eta_i}^{\eta} \!\! d\eta' a^{\prime 3} u(\eta',k) \Bigl\{
\quad \Bigr\} = \frac{a^{\prime 2}}{2 H} u(\eta',k) \Bigl\{\quad\Bigr\}
\Biggl\vert^{\eta}_{\eta_i}\!\! -\frac1{2H} \int_{\eta_i}^{\eta} \!\! d\eta' 
a^{\prime 2} \partial_0' \Bigl[ u(\eta',k) \Bigl\{ \quad \Bigr\} \Bigr] 
\; , } \\
& & = -\frac1{2H} a_i^2 u_i \Bigl\{\quad\Bigr\}_i -\frac1{2H} \int_{\eta_i}^{
\eta} \!\! d\eta' \, a^{\prime 2} \Biggl[\partial_0' u(\eta',k) \Bigl\{\quad
\Bigr\} \!+\! u(\eta',k) \partial_0' \Bigl\{\quad \Bigr\} \Biggr] . \qquad
\label{IBP}
\end{eqnarray}
Of course the term from the lower limit approaches a constant at late times,
and it need not concern us further. The same is true of the integral in 
(\ref{IBP}) which involves the derivative of $u(\eta',k)$,
\begin{equation}
\partial_0' u(\eta',k) = \frac{H}{\sqrt{2 k^3}} \Bigl[-\frac{k^2}{H a'}\Bigr]
\exp\Bigl[\frac{i k}{H a'}\Bigr] \; .
\end{equation}
That leaves only the integral which involves the derivative of the curly
bracketed term of (\ref{exact}),
\begin{equation}
\partial_0' \Bigl\{\quad\Bigr\} = -2k \sin(\alpha) \!\! \int_0^{2\alpha}
\!\!\! dt \, \frac{\sin(t)}{t} \!-\! 2 k \cos(\alpha) \Biggl[
\int_0^{2\alpha} \!\!\! dt \, \frac{\cos(t) \!-\! 1}{t} \!+\! 2
\ln(H \Delta \eta) \!-\! \frac12\Biggr] . \label{dcurly}
\end{equation}

The term we need comes from substituting (\ref{dcurly}) into the final
integral of expression (\ref{IBP}). The contribution from the two $t$
integrals approaches a constant at late times,
\begin{equation}
\frac{k}{H} \int_{\eta_i}^{\eta} \!\!\! d\eta' \, a^{\prime 2} u(\eta',k)
\int_0^{2\alpha} \!\!\! dt \Bigl[\frac{\cos(t \!-\! \alpha) \!-\! 
\cos(\alpha)}{t}\Bigr] \longrightarrow {\rm constant} \; .
\end{equation}
Each of the two remaining terms in (\ref{dcurly}) requires a further 
partial integration to extract the leading late time behavior. For the 
term $k \cos(\alpha)$ we integrate by parts on $a^{\prime 2}$,
\begin{eqnarray}
\lefteqn{\frac{k}{2 H} \int_{\eta_i}^{\eta} \!\!\! d\eta' \, a^{\prime 2} 
u(\eta',k) \cos(\alpha) } \nonumber \\
& & \hspace{1.5cm} = \frac{k}{2 H^2} a' u(\eta',k) \cos(\alpha) \Bigl\vert_{
\eta_i}^{\eta} - \frac{k}{2 H^2} \int_{\eta_i}^{\eta} \!\!\! d\eta' \, a'
\partial_0' \Bigl[ u(\eta',k) \cos(\alpha)\Bigr] \; , \qquad \\
& & \hspace{1.5cm} = \frac{k}{2 H^2} \, a u(\eta,k) + {\rm constant} \; .\qquad 
\end{eqnarray}
The logarithm requires that we partially integrate on $a^{\prime 2} \ln(H
\Delta \eta)$ using the relation,
\begin{equation}
\int \!d\eta' a^{\prime 2} \ln(H \Delta \eta) = -\frac1{H} \Bigl[ (a \!-\! a')
\ln\Bigl(1 \!-\! \frac{a'}{a}\Bigr) \!+\! a' \ln(a')\Bigr] \; .
\end{equation}
The result is,
\begin{equation}
-\frac{2k}{H} \int_{\eta_i}^{\eta} \!\!\! d\eta' \, a^{\prime 2} 
\ln(H \Delta \eta) u(\eta',k) \cos(\alpha) = -\frac{2 k}{H^2} \, a \ln(a) 
u(\eta,k) + {\rm constant} \; .
\end{equation}
It follows that the late time expansion of the final nonlocal term in 
(\ref{calM}) has the form,
\begin{eqnarray}
\lefteqn{-\frac{H^4 a^3}{2^5 \pi^3} e^{-i\vec{k} \cdot \vec{x}} \partial^2 \!\!
\int_{\eta_i}^{\eta} \!\! d\eta' a^{\prime 3} u(\eta',k) \int_{\Delta r \leq 
\Delta \eta} \!\!\!\!\!\!\!\!\!\!\!\!\! d^3x' \, e^{i\vec{k} \cdot \vec{x}'} 
\Biggl\{ \ln\Bigl[\frac14 H^2 (\Delta \eta^2 \!-\! \Delta r^2)\Bigr] \!-\! 
\frac32 \Biggr\} } \nonumber \\
& & \hspace{2cm} = \frac{H^2 a^4}{4 \pi^2} \Biggl\{-\ln(a) u(\eta,k) \!+\! 
\frac14 u(\eta,k) \!+\! O\Bigl(\frac1{a}\Bigr) \Biggr\} . \qquad
\end{eqnarray}

Summing the various local and nonlocal contributions, and recalling that
$\partial_0 u(\eta,k) \sim -k^2/Ha \times u(0,k)$, we find that the
one loop effective mode equation in this gauge takes the form,
\begin{equation}
a^2 \Bigl[ \partial_0^2 \!+\! 2 H a \partial_0 \!+\! k^2\Bigr] \Phi^{\rm non
}_1(\eta,k) = \frac{H^2 a^4}{4 \pi^2} \Biggl\{ -\Bigl[\frac54 \!+\! 48 \pi^2 
\delta \xi^{\rm non}_{\rm fin}\Bigr] u(\eta,k)+O\Bigl(\frac1{a}\Bigr)\Biggr\} .
\end{equation}
To infer the asymptotic solution it is best to change the derivatives from 
conformal time $\eta$ to co-moving time $t \!=\! -\ln(-H\eta)/H$,
\begin{equation}
a^2 \Bigl[\partial_0^2 + 2 H a \partial_0 + k^2\Bigr] = 
a^4 \Bigl[\partial_{t}^2 + 3 H \partial_t + \frac{k^2}{a^2}\Bigr] \; .
\end{equation}
By choosing the finite part of the conformal counterterm appropriately,
\begin{equation}
\delta \xi^{\rm non}_{\rm fin} = -\frac5{12} \frac1{(4 \pi)^2} \; , 
\end{equation}
we reach an equation of the form,
\begin{equation}
\Bigl[\partial_{t}^2 \!+\! 3 H \partial_t \!+\! \frac{k^2}{a^2}\Bigr] 
\Phi^{\rm non}_1(\eta,k) = \frac{H^2}{4 \pi^2} \Biggl\{\frac{C}{a} + 
O\Bigl(\frac{\ln(a)}{a^2}\Bigr) \Biggr\} ,
\end{equation}
where $C$ is a constant. Except for possible homogeneous terms --- which 
can be absorbed into the finite part of the field strength renormalization 
at late times --- the one loop correction rapidly redshifts to zero,
\begin{equation}
\Phi^{\rm non}_1(\eta,k) = -\frac{C}{8 \pi^2} \, \frac1{a} + 
O\Bigl(\frac{\ln(a)}{a^2} \Bigr) \; .
\end{equation}

\section{de Sitter-Lorentz Gauge}

Reliable results also exist for de Sitter-Lorentz gauge \cite{AJ,TW3},
\begin{equation}
\partial_{\mu} \Bigl( \sqrt{-g} g^{\mu\nu} A_{\nu} \Bigr) = 0 \; . 
\end{equation}
Because this gauge preserves manifest de sitter invariance it is best to
express results in terms of the following function of the invariant
length $\ell(x;x')$,
\begin{equation}
y(x;x') = 4 \sin^2\Bigl[\frac12 H \ell(x;x')\Bigr] = a a' H^2 (x \!-\! 
x')^{\mu} (x \!-\! x')^{\nu} \eta_{\mu\nu} \; .
\end{equation}
The imaginary parts appropriate for the polarizations of the Schwinger-Keldysh
formalism are,
\begin{equation}
y_{\scriptscriptstyle ++}(x;x') \equiv a a' H^2 \Delta x^2_{\scriptscriptstyle 
++}(x;x') \qquad {\rm and} \qquad y_{\scriptscriptstyle +-}(x;x') \equiv a a' 
H^2 \Delta x^2_{\scriptscriptstyle +-}(x;x') \; .
\end{equation}
In this notation the scalar self-mass-squared is \cite{PTsW1},
\begin{eqnarray}
\lefteqn{M^2_{{\rm inv} \atop \scriptscriptstyle ++ }(x;x') = - e^2 \delta 
Z^{\rm inv}_{\rm fin} \sqrt{-g} \square \delta^4(x \!-\! x') + 12 e^2 H^2 
\delta \xi^{\rm inv}_{\rm fin} \sqrt{-g} \delta^4(x \!-\! x') } \nonumber \\
& & - \frac{i 3 e^2 H^2}{(4\pi)^4} \sqrt{-g} \sqrt{-g'} 
\square^2 \Biggl\{ \frac{4}{y_{\scriptscriptstyle ++}} \ln\Bigl(\frac{y_{
\scriptscriptstyle ++}}4\Bigr) \Biggr\} + \frac{i 3 e^2 H^4}{(4 \pi)^4} 
\sqrt{-g} \sqrt{-g'} \square \Biggl\{ 7 \Bigl(\frac{4}{y_{\scriptscriptstyle 
++}}\Bigr) \nonumber \\
& & + \Biggl[ 4 \Bigl(\frac{4}{y_{\scriptscriptstyle ++}}\Bigr) - 4 
\ln\Bigl(\frac{y_{\scriptscriptstyle ++}}{4} \Bigr) + 8 \ln\Bigl(1 \!-\! 
\frac{y_{\scriptscriptstyle ++}}4\Bigr) - \frac{6}{1 \!-\! \frac{y_{
\scriptscriptstyle ++}}4} \Biggr] \ln\Bigl(\frac{y_{\scriptscriptstyle ++}}4
\Bigr) \nonumber \\
& & \hspace{7cm} + 8 \sum_{n=1}^{\infty} \frac1{n^2} \Bigl(\frac{y_{
\scriptscriptstyle ++}}4 \Bigr)^n \Biggr\} + O(e^4) \; . \qquad \label{oldinv}
\end{eqnarray}

Before giving the $+-$ term and combining to work out $\mathcal{M}^2_1(x;x')$,
it is best to remove the factors of $1/y$ using the identities,
\begin{eqnarray}
\frac4{y} & = & \frac{\square}{H^2} \Biggl\{ \ln\Bigl(\frac{y}4\Bigr)\Biggr\}
+ 3 \; , \label{1/x} \qquad \\
\frac4{y} \ln\Bigl(\frac{y}4\Bigr) & = & \frac{\square}{H^2} \Biggl\{ \frac12
\ln^2\Bigl(\frac{y}4\Bigr) \!-\! \ln\Bigl(\frac{y}4\Bigr) \Biggr\}
+ 3\ln\Bigl(\frac{y}4\Bigr) - 2 \; . \label{ln/x} \qquad
\end{eqnarray}
Because any analytic function of $y(x;x')$ will cancel in the sum of $++$
and $+-$ polarizations, we do not write out such terms explicitly,
\begin{eqnarray}
\lefteqn{M^2_{{\rm inv} \atop \scriptscriptstyle ++ }(x;x') = - e^2 \delta 
Z^{\rm inv}_{\rm fin} \sqrt{-g} \square \delta^4(x \!-\! x') + 12 e^2 H^2 
\delta \xi^{\rm inv}_{\rm fin} \sqrt{-g} \delta^4(x \!-\! x') } \nonumber \\
& & + \frac{i 3 e^2 H^6}{(4\pi)^4} \sqrt{-g} \sqrt{-g'} \Biggl\{ 
\frac{\square^3}{H^6} \Biggl[ -\frac12 \ln^2\Bigl(\frac{y_{\scriptscriptstyle 
++}}4\Bigr) \!+\! \ln\Bigl(\frac{y_{\scriptscriptstyle ++}}4\Bigr)\Biggr] \!+\!
\frac{\square^2}{H^4} \Biggl[ 2 \ln^2\Bigl(\frac{y_{\scriptscriptstyle ++}}4
\Bigr) \Biggr] \nonumber \\
& & + \frac{\square}{H^2} \Biggl[ -4 \ln^2\Bigl(\frac{y_{\scriptscriptstyle 
++}}4\Bigr) \!+\! 8 \ln\Bigl(1 \!-\! \frac{y_{\scriptscriptstyle ++}}4\Bigr)
\ln\Bigl(\frac{y_{\scriptscriptstyle ++}}4\Bigr) \!+\! 12 \ln\Bigl(\frac{y_{
\scriptscriptstyle ++}}4\Bigr) \!-\! \frac{6 \ln(\frac{y_{\scriptscriptstyle 
++}}4)}{1 \!-\! \frac{y_{\scriptscriptstyle ++}}4} \Biggr] \nonumber \\
& & \hspace{8cm} + {\rm Analytic} \Biggr\} + O(e^4) \; . \qquad \label{inv++}
\end{eqnarray}
The procedure of Section 2 gives the corresponding $+-$ self-mass-squared,
\begin{eqnarray}
\lefteqn{M^2_{{\rm inv} \atop \scriptscriptstyle +- }(x;x') = 
-\frac{i 3 e^2 H^6}{(4\pi)^4} \sqrt{-g} \sqrt{-g'} \Biggl\{ 
\frac{\square^3}{H^6} \Biggl[ -\frac12 \ln^2\Bigl(\frac{y_{\scriptscriptstyle 
+-}}4\Bigr) \!+\! \ln\Bigl(\frac{y_{\scriptscriptstyle +-}}4\Bigr)\Biggr] }
\nonumber \\
& & + \frac{\square^2}{H^4} \Biggl[ 2 \ln^2\Bigl(\frac{y_{\scriptscriptstyle 
+-}}4 \Bigr) \Biggr] \!+\! \frac{\square}{H^2} \Biggl[ -4 \ln^2\Bigl(\frac{
y_{\scriptscriptstyle +-}}4\Bigr) \!+\! 8 \ln\Bigl(1 \!-\! \frac{y_{
\scriptscriptstyle +-}}4\Bigr) \ln\Bigl(\frac{y_{\scriptscriptstyle +-}}4\Bigr)
\nonumber \\
& & \hspace{3cm} + 12 \ln\Bigl(\frac{y_{\scriptscriptstyle +-}}4\Bigr) \!-\! 
\frac{6 \ln(\frac{y_{\scriptscriptstyle +-}}4)}{1 \!-\! \frac{y_{
\scriptscriptstyle +-}}4} \Biggr] \!+\! {\rm Analytic} \Biggr\} + O(e^4) 
\; . \qquad \label{inv+-}
\end{eqnarray}

Of course the delta functions terms in $\mathcal{M}^2_1(x;x')$ make the same 
contributions as they did for the noninvariant gauge of the previous section,
\begin{eqnarray}
\lefteqn{ -\int_{\eta_i}^{0} \!\! d\eta' \!\! \int \!\! d^3x' \Bigl\{-\delta
Z^{\rm inv}_{\rm fin} a^4 \square \delta^4(x \!-\! x') \!+\! 12 H^2 \delta 
\xi^{\rm inv}_{\rm fin} a^4 \delta^4(x \!-\! x') \Bigr\} u(\eta',k) 
e^{i \vec{k} \cdot (\vec{x}' - \vec{x})} } \nonumber \\
& & \hspace{7.5cm} = -12 H^2 \delta \xi_{\rm fin} a^4 u(\eta,k) \; . \qquad
\label{local}
\end{eqnarray}
We can avoid a lengthy computation of the contributions from the nonlocal
terms by taking note of four points:
\begin{itemize}
\item{As explained at the end of Section 2, the only sensible and physically 
interesting regime in which we can solve the effective mode equation is for 
late times, long after the state was released;}
\item{In this regime we can replace the 0th order mode function by a 
constant,\footnote{For $u(\eta',k)$ this is obvious from the fact that 
expression (\ref{uform}) approaches a nonzero constant at late times. That 
the spatial plane wave factor of $e^{i \vec{k} \cdot (\vec{x}' - \vec{x})}$ 
can also be dropped follows from the causality of the Schwinger-Keldysh 
formalism. The factors of $\theta(\Delta \eta \!-\! \Delta r)$ which arise 
whenever $++$ and $+-$ terms are added --- for example, expressions 
(\ref{theta1}-\ref{theta2}) --- require $\Vert \vec{x}' \!-\! \vec{x}\Vert 
\! \leq \! \eta \!-\! \eta'$, which is small in the late time regime.}
\begin{equation}
-\int_{\eta_i}^{0} \!\!\! d\eta' \!\!\! \int \!\! d^3x' \mathcal{M}^2_1(x;x')
u(\eta',k) e^{i \vec{k} \cdot (\vec{x}' - \vec{x})} \rightarrow - u(0,k) \!\!
\int_{\eta_i}^{0} \!\!\! d\eta' \!\!\! \int \!\! d^3x' \mathcal{M}^2_1(x;x')
\, ; \label{replace}
\end{equation}}
\item{This reduces the nonlocal contributions to a sum of terms of the form,
\begin{equation}
-u(0,k) \frac{i 3 H^6 a^4}{(4 \pi)^4} \Bigl(\frac{\square}{H^2}\Bigr)^N \!\!
\int_{\eta_i}^{0} \!\! d\eta' \, a^{\prime 4} \!\! \int \!\! d^3x' 
\Biggl\{ f\Bigl(\frac{y_{\scriptscriptstyle ++}}4\Bigr) \!-\!
f\Bigl(\frac{y_{\scriptscriptstyle +-}}4\Bigr) \Biggr\} \; ; \; {\rm and}
\label{four}
\end{equation}}
\item{The four integrals of the form (\ref{four}) which we require have
been worked out in a previous computation of the vacuum expectation value
of two coincident scalar bilinears at two loop order \cite{PTsW1}.}
\end{itemize}
 
\begin{table}
\vbox{\tabskip=0pt \offinterlineskip
\def\tablerule{\noalign{\hrule}}
\halign to390pt {\strut#& \vrule#\tabskip=1em plus2em&
\hfil#\hfil& \vrule#& \hfil#\hfil& \vrule#\tabskip=0pt\cr
\tablerule
\omit&height4pt&\omit&&\omit&\cr
\omit&height2pt&\omit&&\omit&\cr
&& $\!\!\!\!\! f(x) \!\!\!\!\!$ && $\!\!\!\!\! \int d^4x' a^{\prime 4} 
\{f(\frac{y_{++}}{4}) - f(\frac{y_{+-}}{4})\} \!\!\!\!\!$ & \cr
\omit&height4pt&\omit&&\omit&\cr
\tablerule
\omit&height2pt&\omit&&\omit&\cr
\tablerule
\omit&height2pt&\omit&&\omit&\cr
&& $\!\!\!\!\! \frac{\ln(x)}{1-x} \!\!\!\!\!$ && $\!\!\!\!\!
\frac12 + O(\frac{\ln(a)}{a}) \!\!\!\!\!$ & \cr
\omit&height2pt&\omit&&\omit&\cr
\tablerule
\omit&height2pt&\omit&&\omit&\cr
&& $\!\!\!\!\! {\scriptstyle \ln^2(x) - 2 \ln(1-x) \ln(x)} \!\!\!\!\!$ && 
$\!\!\!\!\! \frac16 - \frac{\pi^2}9 + O(\frac{\ln(a)}{a}) \!\!\!\!\!$ & \cr
\omit&height2pt&\omit&&\omit&\cr
\tablerule
\omit&height2pt&\omit&&\omit&\cr
&& $\!\!\!\!\! {\scriptstyle  \ln(x)} \!\!\!\!\!$ && $\!\!\!\!\! \frac16 
{\scriptstyle \ln(a)} - \frac{11}{36} + O(\frac1{a}) \!\!\!\!\!$ & \cr
\omit&height2pt&\omit&&\omit&\cr
\tablerule
\omit&height2pt&\omit&&\omit&\cr
&& $\!\!\!\!\! {\scriptstyle  \ln^2(x)} \!\!\!\!\!$ && $\!\!\!\!\! \frac16 
{\scriptstyle \ln^2(a)} - \frac89 {\scriptstyle \ln(a)} + \frac74 - 
\frac{\pi^2}{9} + O(\frac{\ln(a)}{a}) \!\!\!\!\!$ & \cr
\omit&height2pt&\omit&&\omit&\cr
\tablerule
\omit&height2pt&\omit&&\omit&\cr
\tablerule}}

\caption{Integrals of de Sitter Invariants. Multiply each term by 
$\frac{i (4 \pi)^2}{H^4}$.}

\label{f4int}

\end{table}

Table~\ref{f4int} reproduces the necessary integrals from previous work
\cite{PTsW1}. Because the integrals can only depend upon $a$, the following 
identity is useful for acting the d`Alembertians,
\begin{equation}
\frac{\square}{H^2} \Bigl(\alpha \ln^2(a) + \beta \ln(a) + \gamma
+ \delta \frac{\ln(a)}{a}\Bigr) = - \alpha \Bigl(6 \ln(a) + 2 \Bigr) 
- 3 \beta + \delta \Bigl[\frac{2 \ln(a) \!-\! 1}{a}\Bigr] \; .
\end{equation}
The triple d`Alembertian nonlocal contribution is,
\begin{eqnarray}
\lefteqn{-u(0,k) \frac{i 3 H^6 a^4}{(4 \pi)^4} \frac{\square^3}{H^6} \!\! 
\int_{\eta_i}^{0} \!\! d\eta' \, a^{\prime 4} \!\! \int \!\! d^3x' 
\Biggl\{ -\frac12 \ln^2\Bigl(\frac{y_{\scriptscriptstyle ++}}4\Bigr) \!+\!
\ln\Bigl(\frac{y_{\scriptscriptstyle ++}}4\Bigr) \!-\! (+-) \Biggr\} }
\nonumber \\
& & \hspace{2.3cm} = u(0,k) \frac{3 H^2 a^4}{16 \pi^2} \frac{\square^3}{H^6}
\left\{ \matrix{-\frac1{12} \ln^2(a) + \frac49 \ln(a) - \frac78 + 
\frac{\pi^2}{18} \cr + \frac16 \ln(a) - \frac{11}{36} + O(\frac{\ln(a)}{a})}
\right\} , \qquad \\
& & \hspace{2.3cm} = u(0,k) \frac{3 H^2 a^4}{16 \pi^2} \Biggl\{ 0 + 
O\Bigl(\frac{\ln(a)}{a}\Bigr) \Biggr\} . \label{non3}
\end{eqnarray}
The double d`Alembertian term gives,
\begin{eqnarray}
\lefteqn{-u(0,k) \frac{i 3 H^6 a^4}{(4 \pi)^4} \frac{\square^2}{H^4} \!\! 
\int_{\eta_i}^{0} \!\! d\eta' \, a^{\prime 4} \!\! \int \!\! d^3x' 
\Biggl\{ 2 \ln^2\Bigl(\frac{y_{\scriptscriptstyle ++}}4\Bigr) \!-\! (+-)
\Biggr\} } \nonumber \\
& & \hspace{.9cm} = u(0,k) \frac{3 H^2 a^4}{16 \pi^2} \frac{\square^2}{H^4}
\Biggl\{ \frac13 \ln^2(a) \!-\! \frac{16}9 \ln(a) \!+\! \frac72 \!-\! 
\frac{2 \pi^2}9 \!+\!  O\Bigl(\frac{\ln(a)}{a}\Bigr) \Biggr\} , \qquad \\
& & \hspace{.9cm} = u(0,k) \frac{3 H^2 a^4}{16 \pi^2} \Biggl\{ 6 + 
O\Bigl(\frac{\ln(a)}{a}\Bigr) \Biggr\} . \label{non2}
\end{eqnarray}
And the term with only a single d`Alembertian is,
\begin{eqnarray}
\lefteqn{-u(0,k) \frac{i 3 H^6 a^4}{(4 \pi)^4} \frac{\square}{H^2} \!\! 
\int_{\eta_i}^{0} \!\!\! d\eta' a^{\prime 4} \!\! \int \!\! d^3x' \!\!
\left\{ \!\!\matrix{-4 \ln^2(\frac{y_{\scriptscriptstyle ++}}4) 
\!+\! 8 \ln(1 \!-\! \frac{y_{\scriptscriptstyle ++}}4) \ln(\frac{y_{
\scriptscriptstyle ++}}4) \!-\! (+-) \cr + 12 \ln(\frac{y_{\scriptscriptstyle 
++}}4) \!-\! 6 \ln(\frac{y_{\scriptscriptstyle ++}}4)/(1 \!-\! 
\frac{y_{\scriptscriptstyle ++}}4) \!-\! (+-) } \!\! \right\} } \nonumber \\
& & \hspace{4cm} = u(0,k) \frac{3 H^2 a^4}{16 \pi^2} \frac{\square}{H^2}
\left\{ \matrix{-\frac23 + \frac{4 \pi^2}9 + O(\frac{\ln(a)}{a}) \cr
+ 2 \ln(a) - \frac{11}3 - 3} \right\} , \qquad \\
& & \hspace{4cm} = u(0,k) \frac{3 H^2 a^4}{16 \pi^2} \Biggl\{ -6 + 
O\Bigl(\frac{\ln(a)}{a}\Bigr) \Biggr\} . \label{non1}
\end{eqnarray}

Summing the local contribution (\ref{local}), and the three nonlocal ones
(\ref{non3}), (\ref{non2}) and (\ref{non1}), gives the following effective
mode equation in Lorentz gauge,
\begin{equation}
a^2 \Bigl[ \partial_0^2 \!+\! 2 H a \partial_0 \!+\! k^2\Bigr] \Phi^{\rm inv
}_1 = \frac{H^2 a^4}{16 \pi^2} \Biggl\{ -12 \pi^2 \delta \xi^{\rm inv}_{\rm 
fin} u(\eta,k) + O\Bigl(\frac{\ln(a)}{a}\Bigr)\Biggr\} . \label{inveqn}
\end{equation}
This is quite similar to the result of the previous section, and the subsequent
analysis is the same. By choosing $\delta \xi^{\rm inv}_{\rm fin} = 0$, and
converting to co-miving time, we obtain the form,
\begin{equation}
\Bigl[\partial_{t}^2 \!+\! 3 H \partial_t \!+\! \frac{k^2}{a^2}\Bigr] 
\Phi^{\rm inv}_1(\eta,k) = \frac{H^2}{16 \pi^2} \Biggl\{C \frac{\ln(a)}{a} + 
O\Bigl(\frac1{a}\Bigr) \Biggr\} ,
\end{equation}
where $C$ is a constant. Except for possible homogeneous terms --- which 
can be absorbed into the finite part of the field strength renormalization 
at late times --- the one loop correction rapidly redshifts to zero,
\begin{equation}
\Phi^{\rm inv}_1(\eta,k) = -\frac{C}{32 \pi^2} \, \frac{\ln(a)}{a} + 
O\Bigl(\frac1{a} \Bigr) \; .
\end{equation}

\section{Discussion}

The Schwinger-Keldysh formalism gives effective field equations which are
causal, and which allow real solutions for Hermitian fields. They are well
adapted to cosmological settings in which the physically sensible experiment
is to release the universe in a prepared state at some finite time. If this
state is chosen to be free vacuum, the formalism will be simple in the sense 
of lacking interaction vertices on the initial value surface. However, the 
unphysical initial state results in effective field equations which diverge
on the initial value surface. Even at late times, there will be some 
contamination from the unphysical initial state in the form of terms which 
decay. This could be avoided by perturbatively correcting the initial state
--- a worthy project which we have, unfortunately, not undertaken. However, 
one can still employ the simple equations reliably at late times where there 
are no divergences and the contamination is dying away.

We have solved the free-vacuum equations for one loop corrections to the 
scalar mode functions of SQED on a locally de Sitter background. Because the
mode functions are gauge dependent, the computation was made in two different 
gauges. In each case it was possible to choose the finite part of the 
conformal counterterm so as to prevent the occurrence of significant 
corrections at late times. 

This might seem a very different outcome from similar studies of one loop
corrections to the mode functions of photons in SQED \cite{PTW2,PW2}, 
fermions in Yukawa theory \cite{PW1,GP} and fermions in Dirac + Einstein
\cite{MW3}. In each of those cases the mode functions suffer one loop
corrections which {\it grow} at late times, instead of falling off. In fact
our result for the scalar mode functions of SQED does fit this pattern for
a generic value of the conformal counterterm. Had we not chosen $\delta 
\xi^{\rm inv}_{\rm fin} = 0$ in (\ref{inveqn}) the one loop correction would
obey the equation,
\begin{equation}
\Bigl[ \partial_t^2 \!+\! 3 H \partial_t \!+\! \frac{k^2}{a^2}\Bigr] 
\Phi_1 = -\frac34 H^2  \delta \xi_{\rm fin} u(0,k) + 
O\Bigl(\frac{\ln(a)}{a}\Bigr) \; .
\end{equation}
In that case the resulting solution would indeed grow at late times,
\begin{equation}
\Phi_1 = -\frac14 \delta \xi_{\rm fin} u(0,k) \ln(a) + 
O\Bigl(\frac{\ln(a)}{a}\Bigr) \; . \label{genxi}
\end{equation}
So the difference between the scalar of SQED and the other cases is just
that SQED possesses a free parameter which can be tuned to suppress the
large one loop corrections that would otherwise occur.

It is worth pursuing this point a little further. Based upon (\ref{genxi}),
the form one expects for the largest late time corrections to the full mode
function is,
\begin{equation}
\Phi(\eta,k) = u(\eta,k) + \sum_{\ell=1}^{\infty} e^{2\ell} \Phi_{\ell}(\eta,k)
\longrightarrow \frac{H}{\sqrt{2 k^3}} \Biggl\{ 1 + \sum_{\ell = 1}^{\infty} 
c_{\ell} \Bigl(e^2 \ln(a)\Bigr)^{\ell} \Biggr\} . \label{asform}
\end{equation}
Secular corrections which involve powers of the infrared logarithm $\ln(a)$
are ubiquitous in quantum field theories that include MMC scalars and/or
gravitons \cite{TW0,TW2,OW1,OW2,SW1,MW3,SW2,PTsW1}. The continued growth of
$\ln(a)$ offers the fascinating prospect of compensating for the small loop
counting parameters --- $e^2$ in this case --- which usually suppress quantum 
loop corrections.
However, the valid conclusion from a series of the form (\ref{asform}) is
that theory breaks down at $\ln(a) \sim 1/e^2$, not that quantum loop effects
necessarily become strong. A nonperturbative analysis is required to 
determine what actually transpires.

Starobinski\u{\i} has developed a stochastic formalism \cite{AAS2} which
has been proven to reproduce the leading infrared logarithms at all orders
in scalar potential models \cite{RPW3,TW5}. Starobinski\u{\i} and Yokoyama
have shown how this formalism can be used to obtain explicit, nonperturbative
results for general expectation values in these models \cite{SY}. 
Starobinski\u{\i}'s formalism has recently been extended to Yukawa theory
\cite{MW2} and, of special interest to us, to SQED \cite{PTsW2,RPW4}. This 
nonperturbative formalism allows us to view the results we have obtained in
a larger context.

First, note that the choice of $\delta \xi^{\rm inv}_{\rm fin} = 0$, which 
enforces the absence of significant late time corrections for de 
Sitter-Lorentz gauge, coincides precisely with the choice which cancels the 
leading infrared logarithm corrections to the expectation value of 
$\varphi^*(x) \varphi(x)$ at order $e^2$ in the same gauge \cite{PTsW1}. This 
same choice cancels the $\varphi^* \varphi$ contribution to the effective 
potential \cite{PTsW2,RPW4}. The concurrence of three distinct results seems to 
confirm the stochastic prediction \cite{PTsW2,RPW4} that significant late time 
effects on the scalar derive solely from the effective potential and not, 
for example, from corrections to the effective action which carry derivatives. 

Our result means that SQED can be renormalized so that the inflationary
production of scalars is not affected at one loop order. That is all we can
conclude from the analysis of this paper, but the stochastic formalism can of
course see to any order. It informs us that the quartic counterterm $\delta 
\lambda$ can be chosen to cancel any $(e^2 \varphi^* \varphi)^2$ contribution 
to the effective potential \cite{PTsW2,RPW4}. This presumably implies that 
$c_2 = 0$ in (\ref{asform}). However, there are unavoidable corrections to the 
effective potential at order $(e^2 \varphi^* \varphi)^3$ and higher, which 
implies that the $c_{\ell}$'s do not vanish for $\ell \geq 3$. So our
result that $c_1 = 0$ is an artifact of being at one loop order which also
happens to be repeated at two loop order. The nonperturbative outcome for 
SQED is that inflationary particle production engenders a nonzero photon 
mass, and the associated vacuum energy prevents the average scalar field 
strength from growing past $\varphi^* \varphi \sim H^2/e^2$ \cite{PTsW2,RPW4}. 
The secular corrections to the scalar mode function (\ref{asform}) represent 
the scalar's initial response to the positive scalar mass associated with the
vacuum energy of massive photons.

\vskip .3cm

\centerline{\bf Acknowledgements}

This work was partially supported by NSF grant PHY-0244714 and by
the Institute for Fundamental Theory at the University of Florida.

\end{document}